\documentclass[
 nopreprint,%choose either nopreprint or preprint
numerical, % this is the default
]{jasatex}

\usepackage{graphicx}% Include figure files
\usepackage{dcolumn}% Align table columns on decimal point
\usepackage{amsmath,amsfonts}% popular packages from the American Mathematical Society
\usepackage{bm}% Bold Math package
\renewcommand{\thefootnote}{\alph{footnote}}

\begin{document}

\preprint{AIP/123-QED}

\title[Giant strain-sensitivity of dissipation]{Giant strain-sensitivity of acoustic energy dissipation in 
solids containing dry and saturated cracks with wavy interfaces}% Force line breaks with \\

\author{V.\,Yu. Zaitsev }
 \email{vyuzai@hydro.appl.sci-nnov.ru}
 %\altaffiliation[Also at ]%
% {Physics Department, XYZ University.}%Lines break automatically or can be forced with \\
\author{L.\,A. Matveev}%

\affiliation{%
Division of Hydrophysics and Hydroacoustics, Institute of  Applied Physics, Russian Academy of Sciences, Uljanova St. 46, Nizhny Novgorod, 603950, Russia}%Line breaks may be forced with \\ here, too

%\affiliation{Second institution and/or address}%

\date{\today}% It is always \today, today,

\begin{abstract}
Mechanisms of acoustic energy dissipation in heterogeneous solids attract much attention in view of their importance for material characterization, nondestructive testing, and geophysics. Due to the progress in measurement techniques in recent years it has been revealed that rocks can demonstrate extremely high strain sensitivity of seismo-acoustic loss. In particular, it has been found that strains of order $10^{-8}$  produced by lunar and solar tides are  capable to cause variations in the seismoacoustic decrement on the order of several percents. Some laboratory data (although obtained for higher frequencies) also indicate the presence of very high dissipative nonlinearity. Conventionally discussed dissipation mechanisms (thermoelastic loss in dry solids, Biot and squirt-type loss in fluid-saturated ones) do not suffice to interpret such data. 
Here,  the dissipation at individual cracks is revised taking into account the influence of wavy asperities  of their surfaces quite typical of real cracks, which can drastically change the 
values of the relaxation frequencies and can result in giant strain sensitivity of the dissipation without  the necessity to assume the presence of unrealistically thin (and, therefore, unrealistically soft) cracks. In particular, these mechanisms suggest interpretation for observations of pronounced amplitude modulation of seismo-acoustic waves by tidal strains.

\end{abstract}

 %, 83.60.Df, 91.60.Lj}% PACS, the Physics and Astronomy
                  % Classification Scheme.83.60.Df Nonlinear viscoelasticity
                 %91.60.LjAcoustic properties (of rocks)
%Use showkeys class option if keyword
\maketitle

\section{\label{sec:level1} Introduction 
%First-level heading:\protect\\ The line
%break was forced \lowercase{via} \textbackslash\textbackslash
}

In recent years, much attention is paid to the so-called mesoscopic 
%(or microstructure-induced) 
nonlinear elasticity\cite{guyer1999} of solids containing such structural features as cracks, contacts, intergrain 
aggregates of dislocations etc. that are small in the scale of the elastic wave length.  
%Indeed, rocks, concretes, many engineering materials, especially with 
%damaged structure, 
%From the viewpoint of their nonlinear acoustic properties such materials  can reasonably be considered as a special and  rather wide class \cite{guyer1999}. 
Quite often the quadratic nonlinear-elastic parameter $\beta$ for such materials can be $10^{3}-10^{4}$ in contrast to $\beta\sim10^0$ typical of ideal crystals of homogeneous amorphous solids. The common feature of the above-mentioned structural features-defects is their very high relative softness compared with that of the surrounding homogeneous material. Thus the local strains at the defects are strongly (often by several orders of magnitude) enhanced. This fact results in strongly increased macroscopic elastic nonlinearity, which can be be instructively elucidated using rheological-level models \cite{zaitsev1996,Belyaeva1997} that are in principle distributed and heterogeneous even if the properties of regions small in the wavelength scale are described.  

In many cases, adhesion or frictional effects are also localized at those soft defects, which 
makes  the resultant nonlinearity hysteretic \cite{guyer1999}. In recent years increasing attention is also paid to 
nonlinear-dissipative properties  of mesoscopic solids, which manifest themselves in rather pronounced
 variations in the dissipation of one elastic wave in the presence of another wave 
 \cite{Zaitsev2000b, Nazarov2002a}  or  an applied (quasi)static stress (although the very fact of pressure-dependence of dissipation or its dependence on the acoustic wave amplitude in rocks has been known for years \cite{Gordon1968, toksoz1979attenuation}). The most striking feature is that quite moderate strains $\varepsilon \sim10^{-6}-10^{-5}$ are able to cause variations in the decrement up to tens of percents or even several times  \cite{Nazarov2002a, Zaitsev2003, Fillinger2006a, Mashinskii2005}, whereas the accompanying variations in the elastic moduli are on the order $\beta\varepsilon$  and do not exceed  $10^{-2}-10^{-3}$. 
 % even for mesoscopic solids with very high nonlinear elastic parameter $\beta\sim10^{3}-10^{4}$.  This means that the relative nonlinearity-induced variations in  the dissipation (i.e., inverse Q-factor or the logarithmic decrement $\theta=\pi/Q$) are  $10^{2}-10^{3}$ times stronger that the simultaneous relative variations in the elastic modulus. 
 
   In addition to laboratory measurements, even more giant strain sensitivity of the dissipation is indicated by some field data. For example, in experiments \cite{glinskii2000vibroseismic} on 
    long-range (357 km and  430 km) propagation of monochromatic elastic wave produced by
     high-stability vibration sources operating at frequencies of $5-7$ Hz, the accuracy of the measurements 
    was sufficient to single out periodic variations in the received-signal parameters which were well-correlated with the periodicity of the lunar-solar tides. For tidal strains in the Earth-crust, the characteristic amplitude \cite{Melchior} is $\sim10^{-8}$ , whereas   in observations \cite{glinskii2000vibroseismic}
     the variations of the received-signal amplitude  amounted to $2-4$ percents and $1-2$ degrees for 
    the  signal phase. These values correspond to the path-averaged \emph{relative} variations in the 
    elastic modulus $\Delta E/E\sim10^{-5}$ and \emph{absolute} variations in the decrement $\Delta\theta\sim3\cdot10^{-5}$ which are of the same order of magnitude. Assuming for the average value of the decrement  $\theta\sim(0.3-1)\cdot10^{-2}$
     (i.e., $Q=\pi/\theta \sim100-300$), we estimate that the 
     \emph{relative} variation in the decrement is $\Delta\theta/\theta\sim0.3\cdot10^{-2}$ and is over two orders of magnitude greater than $\Delta E/E\sim10^{-5}$.  
     
     Taking into account that the tidal strain $\varepsilon_{0}\sim10^{-8}$, we obtain the path-averaged estimate for the quadratic nonlinear parameter
      $\beta=(\Delta E/E)/\varepsilon_{0}\sim 500-700$, which is
      %is significantly greater than for homogeneous solids, but
       not so high as $\beta\sim10^{5}$ reported, for example, in the pioneering observations 
      \cite{de1973solid, reasenberg1974precise} of the tidal variations in the elastic-wave velocities. The above estimated smaller value of $\beta$  
      is not surprising since in the long-range experiments \cite{glinskii2000vibroseismic}  the wave path reached depths of several tens 
      of kilometers where the high pressure of the overburden rock layers closed soft cracks and 
      significantly reduced the path-averaged nonlinearity of the rocks. 
      
      Much higher (comparable with 
      the data \cite{de1973solid, reasenberg1974precise}) values of $\beta$  were observed in cross-well experiments \cite{bogolyubov2004} with a high-stability downhole seismo-acoustic source operating at a frequency of $167$~Hz, the 
      propagation distance $360$~m with the estimated wave velocity along the path about $3000$~m/s. In work \cite{bogolyubov2004} the tide-induced variations for the wave phase were about $0.05$~rad.  and elastic-modulus variations $\Delta E/E\sim10^{-3}$ that indicated the nonlinearity parameter $\beta=(\Delta E/E)/\varepsilon_{0}\sim(1-2)\cdot10^5$ like in works  \cite{de1973solid, reasenberg1974precise}.  For 
      the wave amplitude, its tide-induced variations were about $10$\% and  corresponded to the absolute variations in the decrement  $\Delta\theta\sim(2-5)\cdot10^{-3}$, which means that the strain-sensitivity of the rocks was so giant that the tidal strains $10^{-8}$ were able to produce the relative variations in the decrement  $\Delta\theta/\theta\sim 10^{-2}-10^{-1}$. 
   
  Similar estimates for $\Delta\theta/\theta$, although less directly, are confirmed by the field observations  of the tidal modulation of the intensity of received endogenous seismic noises  at several observation sites at the Kamchatka peninsula and in Japan \cite{Saltykov2006, Saltykov2002}.  Normally, the weak-amplitude 
  noise with strains $10^{-10}-10^{-12}$  was recorded by a sensitive narrow-band receiver around a frequency of $30$~Hz.  Coherent averaging (from several weeks to several months) of the noise envelope made it possible to reliably single out periods
  of individual solar and lunar-tide components in the noise-intensity modulation with a typical depth ranged from $2-3$ 
  to $6-8$\%. Taking into account that for rather weak tidal strains $\varepsilon_{0}\sim10^{-8}$, their direct influence
   on the rock fracturing and the accompanying seismo-acoustic emission does not look very 
  probable, the observed modulation can readily be explained  \cite{Zaitsev2008a} by the tidal modulation of the effective size of the region from which the signal at the receiver is collected. This size is determined by the
   characteristic  damping length for the noise.  Thus, the relative variations in the intensity of the received noise should be 
   proportional to  $\Delta\theta/\theta$. The above 
   obtained estimate $\Delta\theta/\theta \sim10^{-2}-10^{-1}$ based on the independent direct 
   measurements  \cite{bogolyubov2004} 
  % of variations of the propagating-wave amplitude 
well agrees with the depth of the   tidal modulation  of the noise  \cite{Saltykov2002, Saltykov2006}. 

Thus various experimental data require an explanation of the giant value of strain-sensitivity of dissipation in mesoscopic solids. 
  % and, in particular, the explanation of the fact that the 
 %nonlinenarity-induced relative variations in the dissipation $\Delta\theta/\theta$ so strongly exceed %the 
%simultaneous variations in the elastic modulus   $\Delta E/E$. 
Since those strain-induced 
variations in the dissipation are observed for very small amplitudes of probing acoustic waves, for which absolute displacements at the microstructural defects (cracks and contacts) fall into essentially 
sub-atomic range, the responsible mechanism(s) should not involve an activation threshold (unlike hysteretic mechanisms  
of frictional  \cite{Gordon1968, Mavko1979} or adhesion  origin \cite{sharma1994grain}).
 %cannot be applied (we note that this problem of subatomic displacements was already mentioned in \cite{Gordon1968, Mavko1979}). 

As discussed in works \cite{Zaitsev2000b, Zaitsev2006}, pronounced  strain-dependent dissipation in 
mesoscopic solids should arise due to combined action of purely elastic nonlinearity of the soft defects 
and 
 conventional linear (i.e., viscous-like) dissipation that is also localized at the same defects 
because of the locally strongly enhanced strain rate. This mechanism does not 
require a finite threshold (unlike essentially super-atomic displacements required for activation of 
adhesion/frictional phenomena), although it can act in parallel with hysteretic mechanisms. 
   
In what follows, we consider physical realizations of this mechanism that are relevant to solids containing dry and fluid saturated cracks. In both cases,  the dissipation is threshold-less in amplitude: of thermoelastic origin in dry and viscous in fluid-filled cracks.  In both cases the key role is played by the same geometrical features
%that are quite typical 
of real cracks for which corrugated surfaces (having wavy asperities) are typical rather
 than smooth nearly plane-parallel form often used in the dissipation models.    
The analysis will be performed in the style of physical argumentation used by Landau and Lifshitz 
 \cite{Landau-elast}  in the discussion of thermoelastic loss in polycrystalline solids and  in work \cite{Armstrong1980}. Such an 
 asymptotic approach gives clear representation of the physics of the discussed phenomena and ensures quantitative estimates with a reasonable accuracy comparable with that for formally exact solutions obtained for idealized (and thus approximate) 
 models like elliptical cracks, etc. 

\section{ General consequences of  wavy roughness of surfaces in real cracks for
strain sensitivity  of dissipation} 

\begin{figure}

\setlength\fboxsep{10pt}
\setlength\fboxrule{5pt}
\includegraphics[width=8cm]{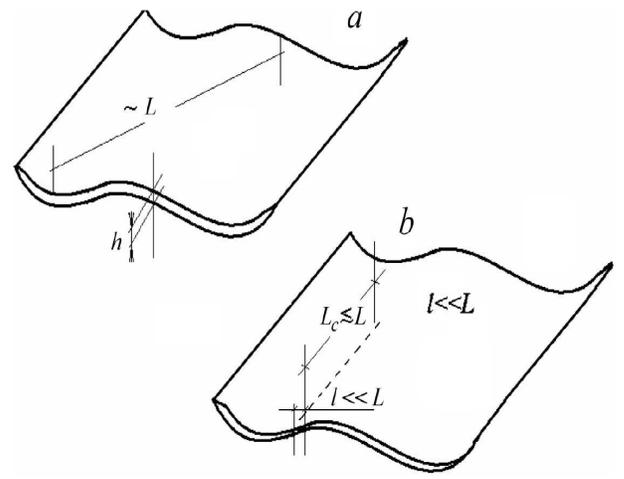}
%\fbox{A real figure would go here}

%\source{\cite{earnshaw1842}}
%\\[0.7cm]
%\vspace{0.7cm}
\caption{
Schematically shown crack with wavy roughness at the interface which results in the creation of elongated (strip-like) contacts or waists inside the crack that may act as a kind of valve for the fluid flow is the crack is liquid-saturated.
}
\label{fig:wavy-crack}
\end{figure}

%In what follows,  we revise the conventionally discussed mechanisms of acoustic loss for both dry and liquid-saturated cracks. 
We emphasize the fact that unlike often assumed near-parallel geometry, for interfaces of real cracks,    wavy forms are quite typical, which is confirmed by crack images obtained by various methods and is in agreement with known models of crack initiation. 
Such  initially coinciding surfaces often are not simply separated in the normal direction but also exhibit certain tangential displacement and create  inside the crack elongated ``waists''  (either nearly contacting or already contacting) as shown in Fig.\,\ref{fig:wavy-crack}.
 Inside liquid-saturated cracks, the so-created narrow waist can act for fluid flows as a kind of 
 valve near which pressure gradients, flow velocities and the corresponding viscous 
 dissipation in the crack should be localized. For dry cracks, the thermo-elastic 
dissipation should also be strongly localized and enhanced at the inner contacts.  In the vicinity of such wavy asperities, the local separation (or 
 interpenetration) $\widetilde{h}$ of the crack surfaces is significantly smaller than the average opening 
 $h$ of the crack. In contrast, the absolute variation in the average opening $h$  and in 
 the local separation $\widetilde{h}$ of the crack surfaces are 
 practically the same, $\Delta h\approx \Delta\widetilde{h}$. Due to this fact, the variations of the contact  pre-strain (or the narrow waist opening) can be $h/\widetilde{h}\gg1$ stronger perturbed than the average opening which determines the loss at the whole crack.   

After finding the acoustic loss at one crack the macroscopic logarithmic decrement $\theta$ can be found as the ratio 
\begin{equation}
\label{eq_def_teta}
 \theta(\omega)=\frac{W_0}{2W_{ac}},
\end{equation}
where the energy  $W_{0}$ is dissipated during one period in a unit volume and $W_{ac}=K\varepsilon^2/2$ is the acoustic energy density stored in the material with the modulus $K$. In the order-of-magnitude estimates, it is not critical to specify which particular modulus is taken. Besides, 
 we also can approximately consider that the presence of cracks does not drastically change the material's elastic modulus $K$.
 %(mostly determined by the homogeneous solid matrix). Therefore,  the density of the elastic-wave energy $W_{ac}$ is weakly dependent on frequency and it is the frequency dependence $W_{0}(\omega)$ which mainly determines the dependence of the decrement $\theta(\omega)$. 

\section{Thermoelastic loss at inner contacts in dry cracks}
Unlike homogeneous materials for which  thermoelastic  dissipation of elastic waves is often negligible,
%The contribution of thermoelastic effects to the dissipation of elastic waves is often considered negligible. This impression is based on the estimates performed for homogeneous materials. However,  
in solids with microstructure  (e.g., polycrystalline \cite{Landau-elast}) thermoelastic dissipation significantly increases due to the presence of small (compared with the elastic-wave length) heterogeneities that strongly increase temperature gradients. 
For crack-containing solids, there is another factor which additionally strongly enhances the thermo-elastic coupling: the stress- and strain concentration at the crack perimeter as considered in work\cite{Savage1966}. For the thermoelastic loss at the entire crack with characteristic diameter $L$, the thermoelastic dissipation exhibits maximum in the vicinity of the relaxation frequency  $f_{L}\approx\kappa/(2\pi \rho C L^{2})$, where $\rho$ is the material density, $C$ the specific heat per unit mass, and $\kappa$ the thermal diffusivity. For  millimeter-size cracks in rocks, this maximum corresponds to frequencies in the  range $10^{-3}..10^{-1}$ Hz. The analysis  \cite{Savage1966} was based on exact solutions for the stress-field distribution near two-dimensional cracks represented as narrow elliptical cavities. However, %to obtain similar results, the use of a particular exact solution is not necessary.  
even  without specifying details of a particular crack model and estimating temperature gradients determined  by the crack size in the low-frequency limit  and by the temperature-wave length in the high-freqeuncy limit, it is possible to evaluate the elastic energy loss using the approach similar to that   used by Landau and Lifshitz \cite{Landau-elast}  for polycrystalline solids.
%as discussed in more detail in works \cite{Zaitsev2002, Fillinger2006}. 

We start from the thermal diffusivity equation for temperature variations
 $\widetilde T$ with respect to the mean value $T_0$
\begin{equation}
\label{eq:Therm-diff}
\frac{\partial \widetilde T}{\partial t}+\frac{\kappa}{C\rho} \Delta  \widetilde T= \gamma
T_0\frac{\partial\ \varepsilon}{\partial t},
\end{equation}
where $\gamma = \mu_{T} K/(\rho C)$ is the Gruneisen parameter of the thermoelastic coupling, $\mu_{T}$ is the thermal expansion coefficient, $K$ is the bulk modulus, $\varepsilon$ is the material dilatation which for the present approximate  consideration can be identified with strain in the field of a compressional wave.
%,  $C$ is the specific heat (per unit mass), $\rho$ is density, and 
%$\kappa$ is the thermal diffusivity. 
The acoustic energy loss due to irreversible heat flows can be found
from the integral\cite{Landau-elast} 
 \begin{equation}
\label{eq:therm-loss-int}
\frac{\partial W }{\partial t} =-\frac{\kappa}{T_0} \int\left(
\bigtriangledown \widetilde T \right)^{2} dV.
\end{equation}
Estimating the gradients in the crack from 
Eq.\,(\ref{eq:Therm-diff}) and evaluating integral (\ref{eq:therm-loss-int}) 
%(see Appendix)
we obtain the following approximate   expressions for the elastic energy dissipated by the crack during one oscillation period, which well agree  with the asymptotic forms of the results presented in  \cite{Savage1966}:
\begin{equation}
\label{eq01}
W^{LF}_{{crack}}\approx2\pi\omega T_{0}(\mu_{T}^{2}K^{2}/\kappa)
L^{5}\varepsilon_{{}}^{2}, \text{ for }\omega\ll\omega_{L}\approx\frac{\kappa
}{\rho CL^{2}},
\end{equation}
\begin{equation}
\label{eq02}
W^{HF}_{{crack}}\approx2\pi T_{0} \frac{\mu_{T}^{2}K^{2}}{\rho C}[\frac{\kappa}{\rho C \omega}]^{1/2}%
L^{2}\varepsilon_{{}}^{2}, \text{ for } \omega\gg\omega_{L},
\end{equation}
\begin{equation}
\label{eq03}
W^{max}_{crack}\approx2\pi T_{0}(\mu_{T}^{2}K^{2}/\rho C)L^{3}%
\varepsilon^{2}, \text{ for } \omega\approx\omega_{L},
\end{equation}
where $\omega$ is the circular frequency and $\omega_{L}$ is the characteristic circular frequency of thermal relaxation determined by the characteristic diameter $L$ of the crack. 

Since we need Eqs. (\ref{eq01})-(\ref{eq03}) only for comparison with similar equations for thermoelastic loss at inner contacts in cracks, here, we only briefly outline their
derivation. First, note that for sufficiently low frequencies, the crack size $L$ is much smaller than both the elastic 
wave length and the  thermal wave length $\lambda_{th}=[\kappa/(C\rho\omega)]^{1/2}$. Therefore, it is the
scale $L$ which determines the gradients of the temperature variations in the elastic-wave field, 
$\bigtriangledown\widetilde{T}\sim\widetilde{T}/L$. Thus in Eq.\,(\ref{eq:Therm-diff}), we can estimate that
$\Delta\widetilde{T} \sim \widetilde{T}/L^2$  and $\partial \widetilde{T}/\partial t \sim \omega \widetilde{T}$. The condition $L \ll \lambda_{th}$ is 
 equivalent to $\omega \ll \omega_L=\kappa/(\rho CL^2)$ and thus $\partial \widetilde{T}/\partial t \ll \Delta\widetilde{T}$, so that the first term in 
 Eq.\,(\ref{eq:Therm-diff}) can be omitted. Since $\partial \varepsilon/\partial t \sim \omega \varepsilon$, 
 Eq.(\ref{eq:Therm-diff}) yields the estimate for the temeprature variation
 $\widetilde{T} \sim\omega \gamma T_0 C\rho L^2/\kappa$. Then we estimate integral (\ref{eq:therm-loss-int})  taking into account that 
the temperature gradient $\widetilde{T}/L$ is localized in the crack vicinity within the characteristic volume $L^3$. Then one obtains $\partial {W}/\partial t \sim \omega^2 {T_0}\gamma^2C^2\rho^{2} L^{5} \varepsilon^{2}$. Finally, for the energy dissipated over one oscillation period $2\pi/\omega$ we recover  Eq.(\ref{eq01}).

	In the high-frequency limit when $\omega \gg \omega_L=\kappa/(\rho CL^2)$, in contrast, the first term $\partial \widetilde{T}/\partial t$ becomes dominant in Eq.\,(\ref{eq:Therm-diff}), so that the 
the temperature variations are adiabatic: 	$\widetilde T\approx \gamma T_{0}\varepsilon$
 and $\lambda_{th}<<L$. In this case, the dissipation is mainly localized in the near vicinity of 
  the crack edge, in which  fairly universal asymptotic
behavior of strain \cite{Kanninen1985} has the form
 $\varepsilon_{cr} \sim \varepsilon(r/L)^{-1/2}$. Here, radius $r$ is counted from the crack edge 
 (and the angular factor in this order-of-magnitude estimate is neglected). When estimating integral 
 (\ref{eq:therm-loss-int}) we have to consider  two regions $r \geq \lambda_{th}$ and   
 $r < \lambda_{th}$. It the latter region, the formally infinite strain $\varepsilon_{cr} \sim \varepsilon(r/L)^{-1/2}$ and strain gradient at
  $r \rightarrow 0$ does not cause divergence of the integral, since the amount of the generated heat is 
  finite and is smeared within the region $r < \lambda_{th}$. As a result, both subregions  $r \geq \lambda_{th}$ and  $r < \lambda_{th}$ give functionally identical and approximately equal
  contributions, so that  $\partial {W}/\partial t \sim \kappa T_{0} \gamma^2 L^2/ \lambda_{th}$ and for the loss over one period we recover Eq.\,(\ref{eq02}). Finally, Eq.\,(\ref{eq03}) can be found as the crossover between Eq.\,(\ref{eq01}) and Eq.\,(\ref{eq02}) at $\omega = \omega_L$.

To estimate the expected relaxation frequency, we can take parameters  $\kappa \sim 0.015$~W/cm/K and $\rho \sim 2.6$ $~g/cm^{3}$  typical of quartz-like rocks and consider cracks with $L \sim 1$ mm,  which yields $f_L=\omega/(2\pi) \sim 10^{-1}$~Hz. 
It has been estimated is
works \cite{Savage1966, Armstrong1980} that thermoelastic losses described by Eqs.\,(\ref{eq01})-(\ref{eq03}) can account  for the observed values of Q-factor in dry rocks, especially for low seismic frequencies.  On the other hand,   
it is seen from Eq.\,(\ref{eq03}) that the height of the relaxation maximum is proportional to the cube of the characteristic crack size $L$. Therefore, to explain the typically  observed in dry rocks values $Q\sim300-1000$  for ultrasonic frequencies  and even for frequencies of the order of $10^2-10^3$ Hz, it is necessary to assume very high densities of very small cracks (of micro-meter scale), which 
%. Such extremely high densities of microscopic cracks 
does not look realistic.       

In this context, the above mentioned narrow inner contacts have much higher relaxation frequencies. However, the question arises whether such contacts (whose volume  is much smaller than that for the entire crack) can dissipate an appreciable amount of the the elastic-wave energy? 
% If the answer is positive, this is especially interesting in the application to low-amplitude dissipation, i.e., for such low strains that are not able to activate the conventionally discussed frictional  \cite{Gordon1968, johnston1979-att-mech} and adhesion-hysteretic \cite{sharma1994grain} loss at contacts in rocks, for which essentially super-atomic displacements are required.  
The expressions for thermoelastic energy loss at the inner contacts in cracks can be obtained much like
Eqs.\,(\ref{eq01})-(\ref{eq03})  taking additionally into account the local concentration of strain at the contacts (see \cite{Zaitsev2002, Fillinger2006}): 
\begin{equation}
\label{eq1}
W^{LF}_{{cont}}\approx2\pi\omega T_{0}(\mu_{T}^{2}K^{2}/\kappa)l^{2}L_{c}
L^{2}\varepsilon_{{}}^{2}, \text{ for }\omega\ll\omega_{l}\approx\frac{\kappa
}{\rho Cl^{2}},
\end{equation}
\begin{equation}
\label{eq2}
W^{HF}_{{cont}}\approx(2\pi/\omega)\kappa T_{0}(\mu_{T}K/C\rho)^{2}L_{c}%
(L/l)^{2}\varepsilon_{{}}^{2}, \text{ for } \omega\gg\omega_{l},
\end{equation}
\begin{equation}
\label{eq3}
W_{cont}^{\max}\approx2\pi T_{0}(\mu_{T}^{2}K^{2}/\rho C)L_{c}L^{2}%
\varepsilon^{2}, \text{ for } \omega\approx\omega_{l},
\end{equation}
 Here, $L_{c}$ in the length of the strip-like contact, $l$ is its width and the other notations are the same as for Eqs.\,(\ref{eq01})--(\ref{eq03}), but the relaxation frequency $\omega_{l}$ is determined by the width $l\ll L$ of the  strip-like contact.
It is seen from  Eqs.\,(\ref{eq1})--(\ref{eq3}) that for frequencies much lower that the relaxation frequency $\omega_{l}$, the loss is growing as a linear function of $\omega$, and for frequencies much greater than  $\omega_{l}$, the loss is inversely proportional to $\omega$. The asymptotic law 
$\omega^{-1}$
is similar to the high-frequency asymptotic for spheroidal voids \cite{Savage1966} rather than flat cracks. 

\begin{figure}
\setlength\fboxsep{10pt}
\setlength\fboxrule{5pt}
\includegraphics[width=8cm]{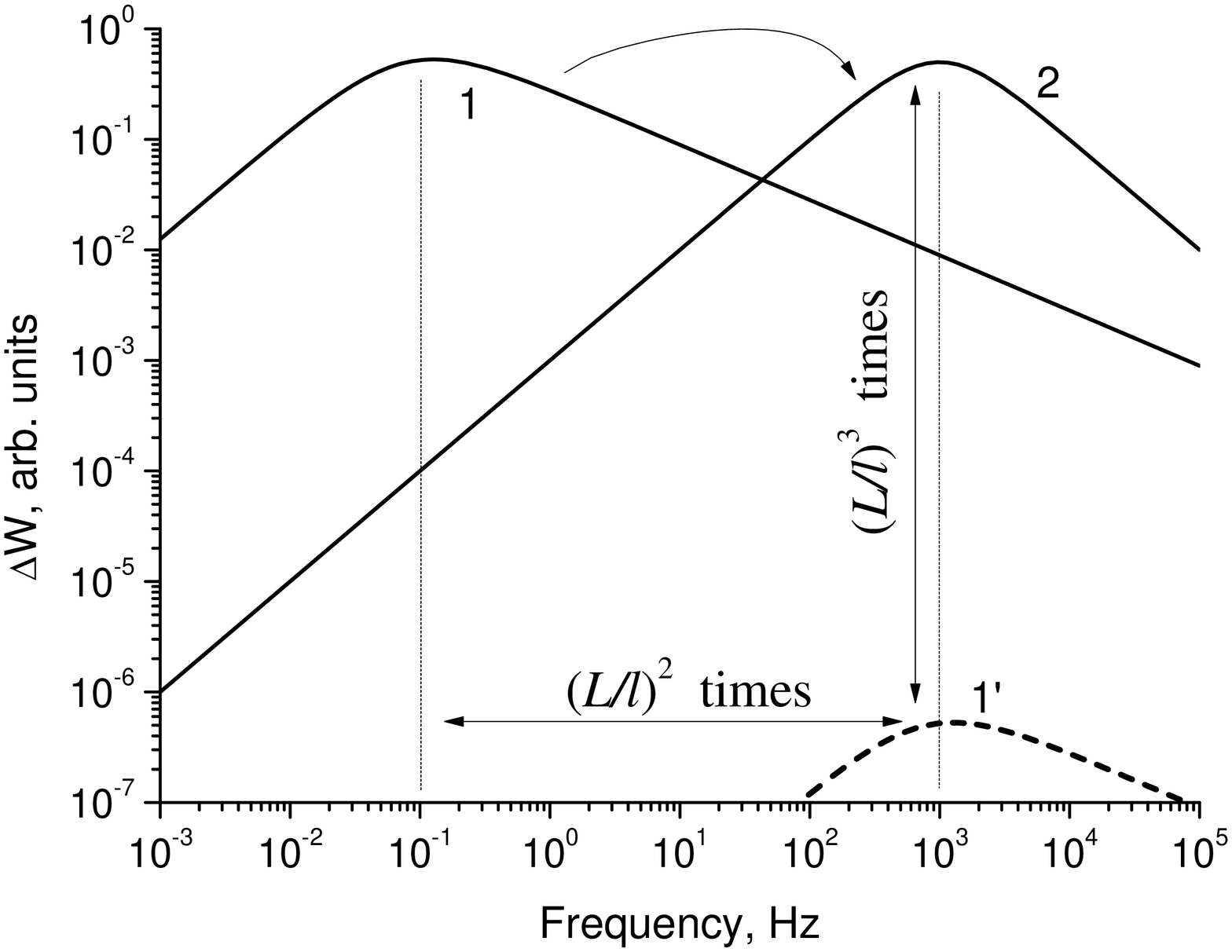}
%\includegraphics[11 8 686 528, width=8cm]{fig-2}
%\fbox{A real figure would go here}

%\source{\cite{earnshaw1842}}
%\\[0.7cm]
\caption{
Schematically shown relative positions and heights of the thermoelastic relaxation peaks for the crack of size $L$ as a whole (curve \emph{1}), for the inner strip-like contact of width $l\ll L$ and length $L_c\sim L$ (curve \emph{2}), and a peak similar to curve \emph{1}, but for a small crack of size $l$ (curve \emph{1'}). It is assumed that $L=1$~mm and $l/L\sim10^{-2}$. 
}
\label{fig:thermo-peaks}
\end{figure}

The most striking conclusion, which is seen from Eqs.\,(\ref{eq1})-(\ref{eq3}),  is  that 
for a strip-like contact with a length $L_{c}\sim L$, the magnitude of loss in the vicinity of the maximum 
located at $\omega_l \gg \omega$ is determined by the size of the whole crack  (since $L_{c} L\sim L^{3}$). Consequently the maximum loss at such  contact is of the same order as for the conventionally considered maximum for the entire crack (compare Eqs.\,(\ref{eq03}) and (\ref{eq3})).  In contrast, the positions of the maxima $\omega_{l} \approx\kappa/(\rho Cl^{2})$ and $\omega_{L} \approx\kappa/(\rho CL^{2})$ on the frequency axis can differ orders of magnitude.  Thus one crack of size $L$ with a strip-like contact $L_c\sim L$ of width $l\ll L$ can produce near the relaxation peak $\omega_l$ the thermoelastic dissipation comparable with the contribution of  $(L/l)^3\gg1$ small cracks with the diameter $l$. Taking again an example of quartz-like rock, this difference is illustrated in Fig.\,\ref{fig:thermo-peaks}  for $L/l=10^2$, which means that \textit{one} such inner contact produces the same dissipation as $10^6$ small cracks of size $l$.     

Noting that  asymptotic Eqs.\,(\ref{eq1})-(\ref{eq3}) agree with the frequency dependence of loss for a classical relaxator
%, one can readily obtain an approximate expression for the loss at the contact in the entire frequency range. Then 
and assuming that the density of such cracks with inner contacts equals $n_{0}$ we obtain from Eq.(\ref{eq_def_teta}) the expression for the decrement in the entire frequency range:
\begin{equation}
\label{eq4}
\theta^{cont}=\frac{2\pi T_{0} \mu_{T}^{2} KL_{c} L^{2} }{\rho C} \frac{\omega
/\omega_{l} }{1+(\omega/\omega_{l} )^{2} } n_{0}
\end{equation}
%We remind that $\theta=\pi/Q$, where $Q$ is the quality factor of the material. 
To find the overall decrement determined by the contributions of contacts with different parameters $L$, $L_{c}$, and $l$ we have to integrate  Eq.\,(\ref{eq4}) over the distribution $n(L,L_{c},l)$, for which it is reasonable to assume that the distribution over the width $l$ should be essentially independent of the distribution over $L$ and $L_{c}$, so that $n(L,L_{c},l)=n(L,L_{c})n(l)$ and the averaging takes the form  
\begin{equation}
\label{eq5}
\theta^{cont}=\frac{2\pi T_{0} \mu_{T}^{2} K}{\rho C} \int  \frac
{L_{c} L^{2}\omega/\omega_{l} }{1+(\omega/\omega_{l} )^{2} } n(L,L_{c} )n(l)dLdL_{c} dl
\end{equation}
We note that a wide distribution $n(l)$ can strongly smoothen the frequency dependence of the dissipation resulting in nearly-constant $Q$-factor even without assuming  very high densities of tiny cracks.  

\begin{figure}

\setlength\fboxsep{10pt}
\setlength\fboxrule{5pt}
\includegraphics [width=8cm] {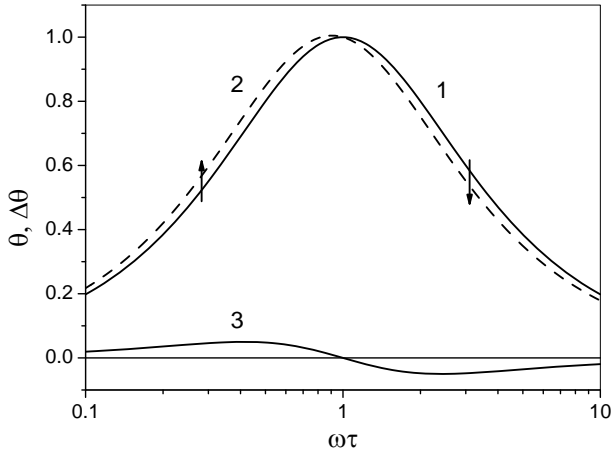}
%\fbox{A real figure would go here}

%\source{\cite{earnshaw1842}}
\caption{
Variation in the shape $\theta(\omega)$ of the normalized relaxation peak for an inner contact (curves \emph{1} and \emph{2}) due to 
15\%  variation in the contact width. The  variation $\Delta \theta$ in the decrement  has opposite signs at the opposite sides from the maximum and peak-to-peak excursion about 15\%  (curve \emph{3}). 
}
\label{fig:peak-shift}
\end{figure}

 It is seen from the structure of integral~(\ref{eq5}) that the height of the relaxation peak and its location on the frequency axis are quite independent since they are determined by essentially different parameters ($L$ and $L_{c}$ for the peak height and $l$ for the position). Evidently, for sufficiently small variation in the average strains and stresses, the width of the contacts can already be significantly perturbed, whereas the length of the contact and the average opening of the crack can be only slightly affected. This can result
 %means that the variation in the relaxation frequency leads to 
 in the displacement of the relaxation maximum on the frequency axis almost without affecting its height as illustrated in Fig.\,\ref{fig:peak-shift} for  the case of identical contacts. 
To estimate how strongly the frequency $\omega_{l} \approx\kappa/(\rho Cl^{2})$ of the maximum of the thermoelastic loss at such a contact  can be perturbed by variation in the mean strain in the material, let us use the presented in Landau and Lifshitz \cite{Landau-elast} solution for the width of the contact between two compressed aligned cylinders: 
\begin{equation}
\label{eq6}
l\approx2\left(  \frac{16DF}{3\pi} \frac{RR^{\prime}}{R+R^{\prime}} \right)
^{1/2},
\end{equation}
where $D=(3/4)[(1-\sigma^{2} )/E+(1-{\sigma^{\prime}}^{2} )/E\rbrack$, $\sigma$  and $\sigma^{\prime}$ 
are Poisson's coefficients of the materials of the cylinders,  $E$ and $E^{\prime}$ are the Young moduli of the cylinders, and $F$ is the specific force per unit length of the contacts. In the case of identical materials with the same accuracy which corresponds to Eqs.(\ref{eq1})--(\ref{eq4}), we can consider that $D\approx 3/(2E)$ and the force $F$ per unit length can be expressed as $F=F_{c}/L_{c}$, where $F_{c}$ is the total force applied to the contact. Then from Eq.\,(\ref{eq6}) we obtain
\begin{equation}
\label{eq7}
l^{2} \approx(16F_{c} R)/(\pi EL_{c} ).
\end{equation}
For the force $F_{c}$ applied to the inner contact  in a crack, an approximate expression $F_{c}/L_{c} \approx\varepsilon EL^{2} /(L+L_{c} )$ can be obtained \cite{Fillinger2006}, where $\varepsilon$ is the mean strain in the material. Then taking into account that $1/2\leq L /(L+L_{c} )\leq1$ one obtains from Eq.\,(\ref{eq6})  the following approximate expressions:
\begin{equation}
\label{eq8}
l^{2}\approx\frac{16}{\pi}\frac{L}{L+L_{c}}\varepsilon RL\approx  \frac{8}{\pi}\varepsilon RL, 
\text{  } \left\vert \frac{\Delta\omega_{l}}{\omega_{l}}\right\vert =\frac{\Delta
(l^{2})}{l^{2}}\approx\Delta\varepsilon\frac{RL}{l^{2}}.%
\end{equation}
This expression clearly elucidates the reason of the extremely high sensitivity of $l^2$ (and, consequently, variations in $\omega_{l}$) with respect to  the variation $\Delta\varepsilon$ in the mean strain. Indeed, the wavy asperities at the crack interfaces often have radius comparable with the characteristic diameter of the entire crack ($R\sim L$), whereas the width of the inner strip-like contact often does not exceed the average opening $h$ of the crack. Therefore, taking into account the above discussed characteristic values of the crack aspect  ratios, we conclude that the factor $RL/l^2$ in Eq.\,(\ref{eq8}) can easily be as great as $10^{6}$--$10^{8}$. This means that even variations in the mean deformation  $\sim10^{-8}$ (which are conventionally considered unable to appreciably influence acoustical parameters of solids) actually can be able to significantly change the width of inner contacts and affect the near-contact acoustic loss (see Fig.\,\ref{fig:peak-shift}). 

 Estimating the overall sensitivity of the thermoelastic dissipation to the mean strain it is necessary to take into account that conventionally considered thermoelastic losses at the crack as a whole \cite{Savage1966} also contribute to the total decrement. 
% This concerns both the fraction of cracks with inner contacts and the other cracks without inner contacts. 
 We  have, however, to recollect  that the characteristic relaxation frequencies  $\omega_{l}\approx\ \kappa/
\rho Cl^{2}$ and $\omega_{L}\approx\ \kappa/
\rho CL^{2}$ are related as $(L/l)^{2}$ and the high-frequency  thermoelastic loss at the whole crack decreases as $(\omega/\omega_{L})^{-1/2}$  (see Eq.(\ref{eq02})) as illustared in Fig.\,\ref{fig:thermo-peaks}. Then near the relaxation maximum $\omega_{l}$  we found that the ratio of the contributions of the loss $\theta_{loc} \left(  \omega=\omega_{l} \right)$ at the inner contacts to  the conventionally considered ``global'' loss $\theta_{glob} \left(  \omega=\omega_{l} \right)$ at the  cracks as a whole has the form
\begin{equation}
\label{ratio-decr}
\frac{\theta_{loc}\left(  \omega=\omega_{l}\right)  }{\theta_{glob}\left(
\omega=\omega_{L}\right)  }\approx\frac{\widetilde{n}_{cr}}{n_{cr}}\left(
\frac{\omega_{l}}{\omega_{L}}\right)  ^{1/2}\approx\frac{\widetilde{n}_{cr}%
}{n_{cr}}\frac{L}{l},%
\end{equation}
where $\widetilde{n}_{cr}$ and ${n}_{cr}$ are the densities of cracks with and without inner contacts,
respectively. Since $L/l\gg1$, it is clear that even a small portion (e.g., a few percents) of cracks with  narrow strip-like inner contacts  can ensure near $ \omega=\omega_{l}$ the same contribution to the decrement as the conventional "global" mechanism of thermoelastic dissipation at whole cracks. Therefore, even quite a small portion of cracks with contacts  can provide the above discussed extremely high sensitivity of the dissipation to small variations in the mean strain in the material. 

%It is also important to emphasize that Eq.\,(\ref{eq3}) shows that the maximum loss at a narrow strip-lke contact is determined by the size of the entire crack (since $L_{c}L^{2}\sim L^{3}$) and  not  by the contact width $l\ll L$, which determines the frequency position $\omega=\omega_{l}$ of the relaxation maximum for the contact. This means that compared with a tiny crack with characteristic dimension $l\ll L$, even one narrow contact in a crack of size $L$ produces $(L/l)^{3}>>1$  times greater contribution to the loss near $\omega=\omega_{l}$. This difference  can be really huge since the factor $(L/l)^{3}$ can easily fall in the range $10^{5}-10^{7}$.  
Even putting aside somewhat exotic manifestations of the giant stress-sensitivity of the dissipation and considering only its mean value, it can be emphasized that in the audible and ultrasonic ranges typical of  laboratory studies, inner contacts  can ensure a fairly strong contribution to dissipation comparable with the  estimates obtained in works \cite{Savage1966, Armstrong1980} for the ``global'' thermoelastic loss at a crack as a whole for lower frequencies, which showed a reasonable agreement with the experimental data.

\section{Viscous loss at fluid-saturated cracks with inner strip-like ``waists'' }

Now let us consider  viscous dissipation in cracks containing liquids. There is  general agreement that along with the dissipation due to ``global'' fluid flows in pore channels (Biot's
mechanism), an important role in the elastic-wave energy dissipation in rocks belongs to the local (or ``squirt'') flows inside cracks \cite{johnston1979-att-mech,mavko1979wave}. Since the geometry of pore channels for the Biot flows rather weakly depends on the average strain and 
stress, it is the squirt-type dissipation in relatively soft narrow cracks which demonstrates much higher 
sensitivity to the variation in the mean strain in the material. In this section we 
%consider this aspect in  detail with the 
focus on very important new features of the squirt mechanism  in the case
 of cracks  with wavy asperities of the interface. 

To estimate the viscous loss in narrow cracks with nearly parallel surfaces we will use the integral expression \cite{landau1987fluid}  for the rate of kinetic energy dissipation in a flow of a viscous fluid between  parallel solid planes:
 \begin{equation}
\label{eq16}
\frac{\partial W_{kin} }{\partial t} =-\frac{\eta}{2} \int\left(
\frac{\partial\upsilon_{x} }{\partial y} \right)  ^{2} dV
\end{equation}
In integral (\ref{eq16}) over the volume of the flow, $Y$ axis is orthogonal to the interface,   $\upsilon_{x}$ is the projection of the liquid-flow velocity on the $X$ axis, and $\eta$ is the fluid viscosity. The integration should be made over the volume of the flow. Using Eq.(\ref{eq16}) we first estimate the amount of energy $W_{0}$ dissipated in a unit volume over one period. 
%$T_{\omega} =2\pi/\omega$ of the acoustic wave with frequency $\omega$, where  $< {\partial W_{kin}/\partial t}>$ is the time-averaged energy dissipation rate. 
%and then using Eq.\,(\ref{eq_def_teta}) 
%find the the logarithmic decrement $\theta$.  
	The form of $W_{0}(\omega)$ can again be understood by using asymptotic estimates very similar to those in the previous section considering now 
%	. However, instead of the thermal-diffusivity equation we will use 
	the linearized Navier-Stokes equation for the flow velocity $\upsilon_{x}$ 
	%which for the considered geometry takes the form
\begin{equation}
\label{eq:Navier}
\rho \frac{\partial \upsilon_x}{\partial t}=-\frac{\partial p}{\partial x}+\eta \frac{\partial^{2} \upsilon_x}{\partial y^2},
\end{equation}
where $p$ is pressure. In view of similar structures of Eq.\,(\ref{eq:Navier}) and Eq.(\ref{eq:Therm-diff}), the roles of the diffusive term and the term with the time derivative can be estimated in a very similar way. 

In order to better delineate the place of our mechanism in the context of conventionally discussed ones we first recall how the well-known properties of the ``global'' Biot loss in fluid-filled channels can be derived from very simple arguments. For low acoustic frequencies, the role of the first inertial term in 
Eq.\,(\ref{eq:Navier}) is negligible, so that 
$|\partial p/\partial x|\approx|\eta \partial^{2} \upsilon_x /{\partial y^2}|$, which corresponds to a fluid  motion as a parabolic
 Poiseuille flow in the channel. Thus the velocity gradient is determined by the 
 entire channel thickness $H$, $\partial^{2} \upsilon_{x}/\partial y^2\sim \upsilon_x/H^2$, and  is frequency independent. The  fluid velocity $\upsilon_{x} $ in the Poiseuille flow is induced by the gradient of the  
 acoustic pressure $p_{ac}$, which  is determined by the acoustic wave length 
 $\lambda_{\omega}\propto \omega^{-1}$, so that 
 $\upsilon_x \propto \partial p/ \partial x \sim p_{ac}/\lambda_{\omega} \propto \omega$.  Taking into account that integral (\ref{eq16}) is evaluated over a frequency-
 independent volume of the flow, the corresponding loss over one period $2\pi/\omega$ is $W_{0} \propto (2\pi/\omega)\cdot\omega^2$, so that $\theta(\omega)_{LF}\propto \omega$.     

With increasing frequency  $\omega$ 
%the viscous wave length $\lambda_{visc}=(2\eta/\rho \omega)^{1/2} It was pointed out in paper \cite{mavko1979wave} that  with increasing$ 
the first inertial term in Eq.\,(\ref{eq:Navier}) increases and becomes comparable with (and even greater than) the viscous term.  This happens when $|\omega \rho \upsilon_x|\approx |\eta\upsilon_x/H^2|$, i.e., for the characteristic frequency  
\begin{equation}
\label{eq:biot-rel-freq}
\omega_c \sim  \eta/(\rho H^2), 
\end{equation}
which is often expressed in a more indirect way via the ratio of porosity to permeability 
instead of the use of the characteristic width of the channels.  The representation of $\omega_c$ in
form  (\ref{eq:biot-rel-freq}) is especially physically  clear and means that the viscous wave length $\lambda_{visc}=(2\eta/\rho \omega)^{1/2}$
becomes comparable with $H$. For $\omega \gg \omega_c $ the velocity gradient becomes localized near the walls within the viscous layer $\lambda_{visc} \ll H$, which 
determines the flow-velocity gradient, and we have $(\partial\upsilon_{x} /{\partial y})^2 \propto (\omega^{1/2})^2 \propto \omega$.  Next, since the last viscous term in  Eq.\,(\ref{eq:Navier}) now can be neglected, we see that the velocity $\upsilon_x$  becomes proportional to the acoustical pressure amplitude $p_{ac}$ without any frequency-dependent
 factor, since  $\partial\upsilon_{x} /{\partial t} \propto \omega \upsilon_{x}$ and   $\partial p/ \partial x \sim p_{ac}/\lambda_{\omega} \propto \omega$ both are proportional to $\omega$. Besides, in integral Eq.\,(\ref{eq16}) the intergration should be made only within the viscous layer with the thickness $\sim \lambda_{visc}\propto \omega^{-1/2}$ (where the gradient is localized) rather than over the entire volume of the fluid flow.
The resultant  frequency dependence of integral (\ref{eq16}) has the form  $ \omega^{-1/2}(\omega^{1/2})^2=\omega^{1/2}$.  The latter estimate should be multiplied by the factor
$2\pi/\omega$ to find the energy loss  $W_0^{HF}$ during one period, so that  for $\omega \gg \omega_c $,  we obtain the asymptotic frequency dependence for 
 the decrement: $\theta(\omega)^{HF}\propto (2\pi/\omega)\omega^{1/2}\propto \omega^{-1/2}$.  
Thus we recovered the well-known asymptotic dependences of the decrement $\theta(\omega)$  
due to
%related to 
the Biot flows in the pore channels. 

Let us now turn to narrow cracks which are much softer objects and can exhibit stronger stress-dependence.
%than the above discussed equant (cylinder-like) pores. 
The above considered arguments based on the Poiseuille-flow approximation in the low-frequency limit remain very similar for flows in cracks and predict the same asymptotic behavior of the decrement $\theta(\omega)^{LF}\propto \omega$.
 As was mentioned in work \cite{mavko1979wave}, the  localization of the shear-flow gradients within a narrow viscous layer $\sim \lambda_{visc}$ for sufficiently high frequencies may also be applied to cracks, but the characteristic frequency is much higher, since the crack opening $h$ is much smaller than the diameter $H$ of the pore channels.  However, in contrast to rigid pore channels, cracks are much softer objects \cite {reasenberg1974precise}, with a characteristic own modulus $\alpha K\ll K$.  For cracks, the aspect ratio $\alpha\sim h/L$ plays the role of their softness parameter: 
% $\zeta \ll 1$ in the above-considered rheological model and 
 due to their planar geometry cracks are roughly $\alpha^{-1}$ times more compliant than the surrounding matrix with respect to both compression and shear. Thus for crack-like compliant pores, another  characteristic frequency of viscous relaxation appears,  which physically corresponds to  the condition that under  sufficiently high-frequency oscillation, the viscous resistance to crack shearing becomes comparable with the elastic response  of the crack to shearing. 
%, which in the rheological model corresponded to $\omega_r=\zeta \Omega=E/(gl)$. In a fluid-filled crack, the fluid flow can be considered incompressible in the first approximation (i.e. assuming 
%$ \mathrm{div}(  \boldsymbol{\upsilon})=0$). Thus   $\upsilon_{x}/L\sim \upsilon_{y}/h$, so that  $\upsilon_{x} \sim 
%\upsilon_{y}L/h\gg\upsilon_{y}$ (i.e., the resultant flow and the viscous force $\eta  {\partial \upsilon_{x}/\partial y\sim \eta\upsilon_{x}/h}$ are nearly tangential). Since $\upsilon_{y} \sim \omega 
%\Delta h$ (where $\Delta h$ is the variation of $h$), the viscous nearly-tangential stress can be estimated as $ \eta\upsilon_{x}/h} \sim \eta  (L/h) \omega \Delta h/h}$. Finally, the normal projection $\sigma^{visc}_{n}$ of the viscous stress is determined by the small angle $h/L$ 
%of the flow to the crack plane, so that  $\sigma^{visc}_{n} \sim \eta  \omega \varepsilon_{cr}$, where $\varepsilon_{cr} =\Delta h/h$ is the strain of the crack. The sought relaxation frequency is obtained by equalling this viscous reaction in 
%the normal direction to the elastic stress  $\sigma^{el}_{n} \sim \alpha K \varepsilon_{cr}$, which yields $\omega_{r} \sim \alpha K/\eta$. This characteristic  frequency (neglecting an insignificant factor on the order of unity) can be identified in work \cite{Walsh1969}, where the contribution of the incompressibility of fluid $K_f$ was additionally included, which should not be done for the liquid squeezed out of the crack in the neighboring pore space as we shall see below.  
 Namely, under tangential displacement $\Delta x$ of the center of the crack, its shear elastic  reaction $\alpha \mu \Delta x/h$ should be compared with the tangential viscous stress $\eta \omega \Delta x/h$. This yields a characteristic frequency 
  \begin{equation}
\label{eq:shear-rel}
 \omega_d \sim \alpha\mu/\eta, 
\end{equation}  
 which was pointed out, for  example, in work  \cite{Walsh1969}.
 %(this is a direct analogue of the relaxation frequency $\zeta \Omega$ in Eq.\,(\ref{Eq:macro}) for  the rheological model for which we had only one dimension). 
 Now we note that for geophysical materials, typically the values of the shear modulus $\mu$ and compression modulus $K$ are of the same order and the corresponding $\omega_d$ is relatively high. 
 %, so that the corresponding  
%relaxation frequencies  fall into MHz and even GHz range. 
For example, for  viscosity $\eta=10^{-3} $ $\mathrm{ Pa}\cdot \mathrm{s}$  typical of water and modulus $\mu \sim10^{10}$ $\mathrm{N/m^2}$ typical of rocks and quite narrow  cracks with $\alpha \sim 10^{-3}-10^{-4}$ we obtain $\omega_d \sim 10^{9}-10^{10}$ rad./s, which evidently can be relevant only to utrasonic laboratory studies, but not to the seismo-acoustic data discussed in the introduction.  

One more characteristic frequency can  be obtained in a similar way, but  considering the viscous resistance of a crack-like defect to the compression in the direction normal to its plane.
This viscous resistance should be compared with the elastic reaction also in the normal direction.   To estimate the normal viscous resistance we have to supplement Eq.\,(\ref{eq:Navier}) with the continuity equation
\begin{equation}
\label{eq:contin}
 \frac{\partial \upsilon_x}{\partial x} + \frac{\partial \upsilon_y}{\partial y} = 0
\end{equation}
since in the first approximation the liquid  can be considered incompressible. For the discussed nearly
Poiseuille flow, $\partial \upsilon_x/\partial x\sim\upsilon_x/L$ and $\partial \upsilon_y/\partial y\sim\upsilon_y/L$, so that from Eq.\,(\ref{eq:contin}) we obtain $\upsilon_x \sim (L/h)\upsilon_y$. Since 
 the variation $\Delta h$ of the crack opening in the acoustic field corresponds to $\upsilon_y\sim \omega \Delta h$, we find that $\upsilon_x \sim (L/h)\omega \Delta h$. Next, in  Eq.\,(\ref{eq:Navier}), we can similarly 
estimate that  $\partial p/\partial x\sim p/L$ and  $\eta \partial^{2} \upsilon_x /{\partial y^2}\sim \eta \upsilon_x /h^2$.  Neglecting the first inertial term in Eq.\,(\ref{eq:Navier}) for the Poiseuille flow and equalling the above estimated terms in the right-hand side of Eq.\,(\ref{eq:Navier}) we obtain the estimate for the 
pressure $p^{visc}$ induced by the viscous flow of the liquid in the narrow gap: 
\begin{equation}
\label{eq:p-visc}
 p^{visc} \sim \eta \upsilon_x (L/h^2) \sim \eta \omega (L/h)^2(\Delta h/h),
\end{equation} 
For the application to real cracks it is important to understand how critical is the parallelism of the crack surfaces (which for real cracks is rather a rule than an exception).  Assuming that the crack opening 
has the values $h_1$ and $h_2$ at the opposite sides we
 conclude that, in the extreme case when at one side the 
 $h_1=0$, the maximal non-papallelism angle is  of the order $h/L \sim\alpha$, where $h$ is the 
 average crack opening. 
 %pressure due to the viscous flow. 
 It is clear that for estimate (\ref{eq:p-visc}), this small non-parallelism is not important. On the other hand, one may argue that if the planes are not exactly parallel  then the tangential viscous force acting along the inclined wall has a component normal to the crack plane.  The viscous stress $\tau^{visc}$ acting along the 
  inclined plane is readily estimated as $\tau^{visc} \sim \eta\upsilon_x/h\sim \eta (L/h) \omega \Delta h/h$, so that its normal component is $\tau^{visc}_n \sim \tau^{visc}\alpha \sim \eta\omega \Delta h/h$. 
 Comparing the latter contribution and Eq.(\ref{eq:p-visc}) for the pressure due to the viscous flow, we see that 
 $p^{visc}$ is $1/\alpha^2>>1$ times greater than $\tau_{n}^{visc}$, so that actually the contribution of the eventual non-parallelism is not significant. Thus we can use Eq.\,(\ref{eq:p-visc}) for the normal viscous reaction even for cracks with not perfectly parallel surfaces.  

Now we have another question: what is the elastic reaction that should be taken for comparison with the above found value\,(\ref{eq:p-visc}) for the viscous normal resistance. In literature different variants  can be found. One variant is by analogy with obtaining Eq.\,(\ref{eq:shear-rel}) to compare Eq.\,(\ref{eq:p-visc}) with the normal elastic reaction of the crack related to its effective modulus, that is, the quantity $\sim \alpha K$ . The latter value corresponds to the well known rule of thumb \cite{Mavko1978} that the effective softness relatively to the matrix material for a crack approximately equals its aspect ratio. This comparison gives us some characteristic frequency often discussed in literature \cite{Murphy1986}
  \begin{equation}
\label{eq:omega_K}
 \omega_K \sim \alpha^3 K/\eta.
\end{equation}
Less formally, it can be said that the viscous flow pressure $p^{visc}$  arisen due to approaching of the crack surfaces by $\Delta h$ (see Eq.(\ref{eq:p-visc})) becomes so high that the near-crack region of the size $L$ (where the elastic stress was initially nearly released) experiences compression by the value $\Delta h^{*} \sim p^{visc}/(KL)$. For $\omega \sim \omega_K$, the value of $\Delta h^{*}$ becomes comparable with the initial approaching $\Delta h$ and strongly compensates the latter.     
% (and by the way, a similar comparison of with $\tau^{visc}_n$ found for non-parallel wall formally yields another parameter combination with the dimension of frequency $\omega \sim \alpha K/\eta$).
  If the crack is ``instantaneously'' compressed, initially the liquid remains almost "frozen" and then is redistributed during the characteristic time
  $1/\omega_K$ when the elastic compression $\Delta h^{*}$ of the near-crack region gradually releases. For sufficiently high frequency $\omega \gg \omega_K$, the actually 
  displaced volume of the liquid is equal to a small fraction $\omega_K/\omega$ of the volume that would be 
  quasistatically displaced, 
  %for $\omega \ll \omega_K$
 so that if the flow structure remains of the Poiseuille type, the amount of the dissipated energy during one period decreases as $(\omega/\omega_K)^{-1}$. This observation will be used below to estimate the absorption in the relaxation maximum.

 It can also be pointed out that besides the relaxation frequency (\ref{eq:omega_K}) another relaxation frequency for cracks filled by compressible fluids is also discussed \cite{johnston1979-att-mech}. Namely, in view of the fact that modulus $K_f$  of the filling fluid typically is at least an order of magnitude smaller than modulus $K$, it can be argued that the surrounding rock in the first approximation compresses the fluid as ``absolutely rigid'' body. Thus comparing the pressure of the compressed 
  fluid $K_f \Delta h/h$ with the above  found normal viscous reaction (\ref{eq:p-visc}) of the fluid flow, 
  we obtain another characteristic frequency $\omega_r \sim K_f \alpha^2/\eta$ which is determined by the compressibility of the fluid and in contrast to Eq.(\ref{eq:omega_K}) does not depend on the modulus of the rock.%\footnote{#1} %
  \footnotemark[1]  
  This new relaxation frequency differs from $\omega_K$ given by
  Eq.(\ref{eq:omega_K}) by a factor of $K_f/(K\alpha)$. For narrow cracks with $\alpha \ll 1$ and  typical liquids like water or oil,  for which $K_f/K\sim 10^{-1}$, the  factor  $K_f/(K\alpha)$ is quite large, so that
   $\omega_r \gg \omega_K$. In such a case, 
  long before the frequency $\omega_r$ would be attained, the fluid flow should already be strongly damped for $\omega > \omega_K$ because of the finite rigidity of the near-crack region as discussed above. Therefore, the initial low-frequency asymptotic $\propto \omega$ cannot be extrapolated till the 
  frequency $\omega_r \gg \omega_K$. Consequently, the new (and potentially stronger) relaxation maximum near 
   $\omega_r$ simply will not be formed for typical liquids (like water and oil), for which  $K_f/(K\alpha)\gg1$.  Only for sufficiently high-compressible (gaseous) fluids with $K_f/(K\alpha)<1$ this mechanism should form the relaxation maximum near $\omega_r$. But in this case the relaxation frequency $\omega_r $ should necessarily lie lower than $\omega_K$.    
 %For  viscosity $\eta=10^{-3} $\mathrm{ Pa}\cdot \mathrm{s}}$  and the bulk modulus $K_{f}=2.25\cdot10^9$ typical of water  $\omega_r \sim \alpha^2\cdot10^{12}$~rad/s. 
  
%Therefore,  in comparison with Eq.(\ref{eq:shear-rel}) for the shear viscous relaxation (even putting aside the difference between the moduli $\mu$ and $K_f$), the considered relaxation peak can additionally be shifted by a factor of $\alpha<<1$ towards lower frequencies, although very small $\alpha \sim10^{-4}$ are required to reach kHz frequencies, whereas seismic frequencies $10^0-10^2$ Hz require even thinner cracks. 
Less formally, the physical meaning of this type of relaxation is that
%, even if the size of the fluid volume is much smaller than the acoustic wave length,   
a finite time is required for a compressible fluid  
squeezed inside a crack to form the flow which redistributes the fluid end equalizes the internal 
pressure. 
%However, it should be emphasized that for the considered nearly uniform thickness of fluid flow and for
%``normal'' liquids like water or oil (rather than gases), the relaxation peak should be determined by the finite compressibility of rock in the crack vicinity rather than the compressibility of the liquid. 

In any case, the character of such flows can noticeably be perturbed only if the mean opening of the entire 
crack  is perturbed. This means that the mean perturbing strain should already be comparable with 
$\alpha$ 
%. Therefore, 
like for the above-considered ``global'' thermoelastic loss. 
%,  the "global" viscous loss in the crack
 %as a whole yet cannot ensure sufficiently high sensitivity of the dissipation to explain the influence of tidal strains on the dissipation as discussed in the introduction.
 
  \footnotetext[1]
{Finally, another type of the characteristic frequency related to the fluid compressibility was considered in paper \cite{mavko1979wave}. For this characteristic  frequency, the elastic wave length becomes comparable with the length of the liquid drop inside the crack, so that the liquid drop cannot be already  considered as a lumped object in the scale of the elastic wave length. If the liquid occupies almost all crack of diameter $L$, this frequency is given by the condition $\omega \sim (K_{f}/\rho)^{1/2}/L$, which corresponds to fairly high ultrasonic frequencies and will not be further discussed.}

%Besides, the physical meaning of this  transition between lumped and distributed cases is not a relaxation phenomenon in the proper sense. 
  In what follows, we consider  a modified form of the viscous loss in cracks with wavy roughness of the  
surface. This mechanism relates to the conventionally considered viscous
  loss at cracks much like the thermoelastic loss at the inner contacts relate to the thermoelastic loss 
  at a crack as a whole. Even for ``normal'' liquids, the new strongly shifted to lower frequencies and very intense  relaxation maximum is related to the fluid compressibility rather than  to compressibility of rock in the crack vicinity. Besides, which is especially interesting, the stress sensitivity of such a mechanism can  be additionally enhanced orders of magnitude and  can readily explain the experimental data discussed in the introduction.   
  %We will see that under some conditions, 
 %Even for ``normal'' liquids ,
%the relaxation maximum can be formed due to the fluid compressibility rather than due to compressibility of rock in the crack vicinity.  Such relaxation peaks can be formed for
%rather low frequencies corresponding to seismoacoustic range and the intensity of such
%peaks can be really huge. 
 %The key issue for the modified 
%relaxation mechanism is again the wavy roughness of the crack 
%surface, which introduces radical qualitative and quantitative changes in the relaxation character. 
  
%Namely, the characteristic frequency can additionally be strongly shifted down to units of Hz even for  cracks that are in average only moderately narrow. Besides, which is especially interesting, the stress sensitivity of such a mechanism can  be additionally enhanced 2-3 orders of magnitude and  can easily explain the experimental data discussed in the introduction.     

\begin{figure}
\setlength\fboxsep{10pt}
\setlength\fboxrule{5pt}
\includegraphics [width=8cm] {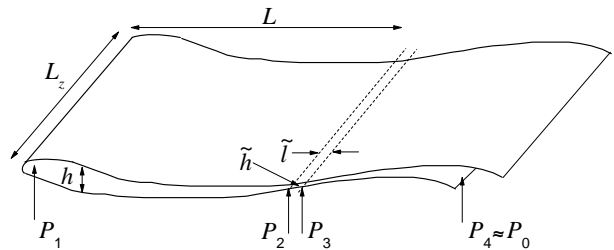}
%\includegraphics [3 14 821 342, width=8cm] {fig-4}
%\fbox{A real figure would go here}
%\source{\cite{earnshaw1842}}
\caption{Schematically shown crack with a narrow strip-like waist created by wavy asperities. The dominant part of the total difference in the pressure $|P_{1}-P_{4}|$ in the fluid flow is localized near the narrow waist, so that $|P_{1}-P_{4}|/|P_{3}-P_{2}|\approx1$. $P_{4}$ is close to the equilibrium pressure $P_{0}$ in the outer pore channel.}
\label{fig:waist}
\end{figure}

Consider a crack with a 
``waist'' created by wavy asperities as shown in Fig.\,\ref{fig:waist}. The notations used are clear from the figure. Here, 
quantities $P_i$ characterize the pressure in the respective cross sections of the crack. Assuming again the Poiseuille character of the flow for sufficiently low frequencies, the 
nearly plane fluid  flow $q=L_z\int\upsilon_x(y)dy$ in the main part of the crack and in the narrower region of the waist should be equal:
\begin{equation}
q=\frac{h^3L_z}{12\eta}\frac{P_2-P_1}{L}=\frac{\tilde{h}^3L_z}{12\eta}\frac{P_3-P_2}{\tilde{l}},
\label{eq:flow1}
\end{equation}
where the meaning of the pressure gradients ${(P_2-P_1)}/{L}$ and ${(P_3-P_2)}/{\tilde{l}}$ is clear
 from  Fig.\,\ref{fig:waist}. Then Eq.\,(\ref{eq:flow1}) yields 
 \begin{equation}
\frac{P_2-P_1}{P_3-P_2} \sim \frac{\tilde{h}^3}{h^3}\frac{L}{\tilde{l}},
\label{eq:flow2}
\end{equation}
so that for sufficiently narrow waist in the crack,   $(\tilde{h}^3/{h^3})({L}/ \tilde{l}) \ll 1$, we conclude 
that 
$(P_2-P_1) \ll (P_3-P_2) \approx \Delta P_{tot}$, that is, the total drop $\Delta P_{tot}$ of pressure is 
mainly localized in the region of the 
narrow waist.  Now imagine an instantaneous slight compression of the crack under an external action 
after which the fluid compressed inside the crack flows through the narrowing into the outer region to 
equalize the pressure. This
flow will exponentially decrease with a characteristic relaxation time~$\tau$. Consequently the total 
volume $q_{\tau}$ 
of the leaked fluid during this relaxation process is given by the integral
  \begin{equation} 
  q_{\tau} = q\int_0^\infty exp(-t/\tau)dt =q\tau,
\label{eq:flow3}
\end{equation}
where $q$ is the flow value corresponding to the initial excess of pressure inside the crack over the pressure in the outer space. Since the rigidity of the outer pore space is much greater than that of the thin crack, the change in the volume of this rigid pore fraction is negligibly small compared with the strain of the crack. Consequently, the outer pore pressure remains close to its equilibrium value $P_0$ and we can consider that $P_3-P_2\approx P_1-P_4\approx P_1-P_0$. Thus the initial value of the flow
\begin{equation}
q=\frac{h^3L_z}{12\eta}\frac{P_3-P_2}{\tilde{l}}\approx \frac{\tilde{h}^3L_z}{12\eta}\frac{P_1-P_0}{\tilde{l}},
\label{eq:flow4}
\end{equation}

Since the variation $\chi$ in the volume of the liquid and the pressure $P$ are related via the bulk modulus $K_f$ of the liquid, $P=\chi K_{f}$, then the total  liquid volume $ q_{\tau} =q\tau$ displaced in order to equalize the pressure can be written via the initial volume $C_0$ of the fluid-filled crack as 
\begin{equation}
q_{\tau}=C_0(\chi_1-\chi_0)=C_0K_f{(P_1-P_0)},
\label{eq:flow5}
\end{equation}
Equations (\ref{eq:flow4}) and (\ref{eq:flow5}) yield
\begin{equation}
q_{\tau}=\frac{h^3L_z}{12\eta}\frac{K_f(\chi_1-\chi_0)}{\tilde{l}}\tau \approx C_0{(\chi_1-\chi_0)}.
\label{eq:flow6}
\end{equation}
Assuming that the sizes of cracks along X and Z directions are of the same order $L_z \sim L$, the crack volume
can be estimated as $C_0\approx hL^2$, so that from Eq.\,(\ref{eq:flow6}) we readily obtain the relaxation time $\widetilde \tau$ and
its inverse value, i.e., the relaxation frequency $\widetilde{\omega}_r$ (tilde denotes that we consider the crack with a narrowing):
\begin{equation}
{\widetilde{\omega}_r}=\frac{\alpha^2 K_f}{12\eta} \left( \frac{\tilde{h}}{{h}} \right)^3 \frac{L}{\tilde{l}}, \text{ and   }
{\tilde{\tau}}=\frac{12\eta}{\alpha^2 K_f} \left( \frac{h}{\tilde{h}} \right)^3 \frac{\tilde{l}}{L}.
\label{eq:flow7}
\end{equation}
 Here, we intentionally singled out the squared aspect ratio $\alpha^2$ for convenience of comparison with the crack without waists  \cite{johnston1979-att-mech}. The absence of the narrow waist can be interpreted as $\tilde{l}\sim L$ and 
 $\tilde{h}\sim h$, i.e.,   $(\tilde{h}/ h)^3\tilde{l}/L \sim 1$ so that Eq.\,(\ref{eq:flow7}) reduces to  
\begin{equation}
{{\omega}_r}=\frac{\alpha^2 K_f}{12\eta}, \text{ and   }
{\tau} \approx\frac{12\eta}{\alpha^2 K_f}.
\label{eq:flow8}
\end{equation}
Within the accuracy of its derivation Eq.\,(\ref{eq:flow8}) coincides with the result $\tau\approx {8\eta}/{\alpha^2 K_f}$ obtained in work 
\cite{johnston1979-att-mech} for narrow cracks without the waist. 

The relaxation times in Eqs.\,(\ref{eq:flow7}) and (\ref{eq:flow8}) do not depend on the elastic modulus of the solid matrix. However, we have already mentioned that for cracks with uniform opening, the relaxation peak (\ref{eq:flow8}) can actually be formed only for very highly compressible fluids (i.e., gases rather than liquids), for which $K_f/K \leq \alpha$, otherwise the compressibility of rock in the 
crack vicinity should dominate and form the relaxation peak near $\omega_K$ given by Eq.(\ref{eq:omega_K}). 

 In contrast to this, the presence of the narrow waist creates a strong obstacle for the fluid leaking.
Therefore, for a given frequency,  this increases the role of the liquid compression, whereas 
the pressure $p^{visc}$ and the deformation of the rock in the crack vicinity is strongly reduced. All this results in a very strong reduction of the relaxation frequency $\widetilde{\omega}_r$  
related to the compressibility of the fluid
 (compare Eq.\,(\ref{eq:flow8}) and Eq.\,(\ref{eq:flow7}) containing the additional small factor $(\tilde{h}/ h)^3\tilde{l}/L \ll 1$). Thus even for ``normal'' liquids like water and oil, the frequency 
 $\widetilde{\omega}_r$ of the new relaxation maximum  becomes comparable and even lower than $\omega_K$, that is, shifts from ultrasonic frequencies to  the  seismoacoustic range directly relevant to the experiments \cite{glinskii2000vibroseismic,bogolyubov2004,reasenberg1974precise,de1973solid,Saltykov2006} discussed in the Introduction.  

 %Since the local opening $\tilde{h}$ is much stronger affected by the variation in the crack opening It is also very important that even in comparison with rather high sensitivity of the 

 Compared with the mean crack opening $h$, the local opening $\tilde{h}$ of the waist is much more sensitive  to variations in the mean strain in the medium. 
 %, sensitivity of the local opening $\tilde{h}$ of the narrow waist is even much higher. This fact results in additionally 2-3 orders higher sensitivity of $\widetilde\omega_r$ given by
%Eq.\,(\ref{eq:flow7}) compared with conventionally discussed $\omega_r$ given by Eq.(\ref{eq:flow8}). 
Therefore 
%the variation in the local opening is sufficient to significantly influence the viscous loss in such a crack and
 it is not necessary to assume the existence of unrealistically thin cracks to ensure the same sensitivity for the mean opening of the crack. The conditions ensuring the Poiseuille type of the flow are even better fulfilled in the vicinity of the waist, so that 
%Since the relaxation frequencies for the discussed mechanism are fairly low and the local 
%	opening near the waist is even much smaller than the average one,  we can be assured that 
%	the local opening is  smaller than the viscous wave length and the liquid flow remains of Poiseuille type. Thus 
	we can apply the considered in the beginning of this section aguments concerning the low- and high-frequency asymptotic behavior of the decrement (proportional to 
	$\omega$ and $1/\omega$, respectively).  This means that the dissipation due to the discussed mechanism can be well approximated by the frequency dependence for a standard relaxator:  
\begin{equation}
\label{eq13}
\theta \approx\theta_0 \frac{\omega/\widetilde{\omega}_{r} }{1+(\omega/\widetilde{\omega}_{r} )^{2} },
\end{equation}
where the relaxation frequency $\widetilde{\omega}_r$ is given by Eq.\,(\ref{eq:flow7})   for a crack with a narrow waist.

% and Eq.\,(\ref{eq:flow8}) for a crack without the waist.

%We note that a rather complex equation for the dissipation presented in paper  \cite{johnston1979-att-mech} 
%is also well approximated by the dependence Eq.\,(\ref{eq13}) 
%in the vicinity of the relaxation maximum corresponding to Eq.\,(\ref{eq:flow8}) for narrow cracks without the waist . 

%Now after estimating the relaxation frequency and the general frequency dependence for the considered disspation mechanism we have to analyze now the variation in the local opening of the waist in the crack influences the position of the relaxation maximum $\widetilde{\omega}_{r}$ and value of the viscous decrement. 

%$\theta$ itself, it is not necessary to assume the existence of crack with unrealistically small aspect ratio in order to ensure the same strain-sensitivity of the dissipation. 

Now, using Eq.\,(\ref{eq16})
%to understand how intense is the discussed relaxation maximum and to analyze more accurately how the relaxation curve is deformed under the influence of the mean strain in the material, we have to find the explicit form of the factor $\theta_0$ in Eq.\,(\ref{eq13}). We 
we will determine 
the prefactor $\theta_0$ in two cases: for a crack with the narrow waist and for a similar in size 
crack with uniform opening, although in the latter case the formation of the relaxation peak $\omega_r$ can 
be possible only if $K_f/K \leq \alpha$, that is, for gaseous fluids (or for unrealistically rigid solid matrix in the case of ``normal'' liquids). Nevertheless, such an
expression will be useful for comparison. 
%The calculation will be based on Eq.\,(\ref{eq16}) for the kinetic energy loss in a two-dimensional  flow of a viscous fluid between two parallel solid planes.

The velocity profile for the Poiseuille's flow in the crack with uniform opening $h$ has a parabolic 
form
 \begin{equation}
\label{eq16A}
\upsilon_x=-\frac{1}{2\eta}\frac{\partial P }{\partial x}y(h-y) =-4\upsilon_{\max}\frac{y(h-y)}{h^2},
\end{equation}
where $\upsilon_{max}=-\frac{1}{8\eta}|\frac{\partial P }{\partial x}|h^2$ is the maximal 
velocity of the fluid in the flow.  Similarly, for the narrow  
waist, where the losses and gradients are localized, we have  
 \begin{equation}
\label{eq16B}
\tilde{\upsilon}_x=-\frac{1}{2\eta}\frac{\partial P }{\partial x}y(\widetilde{h}-y) =-4\widetilde{\upsilon}_{\max}\frac{y(\widetilde{h}-y)}{\widetilde{h}^2},
\end{equation}
where $\widetilde{\upsilon}_{max}=-\frac{1}{8\eta}|\frac{\partial P }{\partial x}|h^2$. 
Equations (\ref{eq16}), (\ref{eq16A}), (\ref{eq16B}) yield for $\partial W_{kin} /\partial t$ 
the following expressions in the discussed two cases of the crack without  and with the waist: 
\begin{equation}
\label{eq17}
\frac{\partial W_{kin} }{\partial t} =-\frac{8}{3} \eta\upsilon_{\max}^{2}
\frac{L_{x} L_{z} }{h}
\end{equation}
\begin{equation}
\label{eq18}
\frac{\partial \widetilde{W}_{kin} }{\partial t} =-\frac{8}{3} \eta\widetilde
{\upsilon} _{\max}^{2} \frac{\widetilde{l} L_{z} }{\widetilde{h} }
\end{equation}
In the estimates below we will assume that $L_{x}\sim L_{z}\sim L$.

Let us now relate the characteristics of the flow with the mean strain in the crack-containing solid. When the mean 
macroscopic strain $\varepsilon$ is created in the material, the crack volume correspondingly changes, which creates the 
flow of the liquid inside the crack.  We recall \cite{Mavko1978} that the crack is a soft object whose relative (compared with the 
surrounding matrix) compliance $\zeta$ is approximately equal to the crack aspect ratio, $\zeta \sim \alpha \sim h/L$. Therefore, the own strain $\varepsilon_{cr}$
of the crack (i.e., the relative variation in its volume) can be estimated as 
$\varepsilon_{cr} \approx\varepsilon/\alpha \approx \varepsilon L/h$, where $L$ is the characteristic size of the crack,
 so that its volume can be estimated as $V_{cr}\sim L^{2} h$. Then for the variation in the crack volume $\Delta V_{cr}$  one obtains $\Delta V_{cr} \approx\varepsilon_{cr} V_{cr} \sim \varepsilon L^{2}
h\cdot(L/h)=\varepsilon L^{3}$. 
Under sinusoidal  oscillatory variation of the crack volume with frequency $\omega$ and amplitude $\Delta V_{cr}$, we can readily relate the rate $\omega\Delta V_{cr}$ of the crack-volume variation with the maximum velocity of the Poiseuille's flow of the fluid inside the cracks without and with the waist:
\begin{equation}
\label{eq19}
\upsilon_{max} = \frac{3}{2}  \frac{ \omega \varepsilon L^{2} } {h}
\end{equation}
\begin{equation}
\label{eq20}
\widetilde{\upsilon} _{max} = \frac{3}{2} \frac{ \omega \varepsilon L^{2} } {\widetilde{h}}
\end{equation}

Substituting these expressions into Eqs.(\ref{eq17}) and (\ref{eq18}) for cracks without and with the waist we obtain 
%the following formulas 
for the period-averaged amounts of the dissipated energy expressions via the amplitude of the mean strain $\varepsilon$:
\begin{equation}
\label{eq21}
\left(  \frac{\partial W_{kin} }{\partial t} \right)  _{aver} =-3\eta
\omega^{2} \varepsilon^{2} \frac{L^{6} }{h^{3} }
\end{equation}

\begin{equation}
\label{eq22}
\left(  \frac{\partial\widetilde{W} _{kin} }{\partial t} \right)  _{aver}
=-3\eta\omega^{2} \varepsilon^{2} \frac{L^{5} \widetilde{l} }{\widetilde{h} ^{3} }
\end{equation}

In Eqs.\,(\ref{eq21}) and (\ref{eq22}) we took into account that for sinusoidal strain with amplitude $\varepsilon$, the period-averaged value of its square equals $\varepsilon^{2}/2$.  We also take into account that the energy loss during one period $T=2\pi/\omega$ is $W_{0} =(2\pi/\omega)(\partial W_{kin} /\partial t)_{aver}$ and the density of the accumulated elastic energy is $W_{el} \approx K\varepsilon^{2} /2$ (since in the used approximation we assume that the elastic modulus does not significantly change due to the presence of the cracks).  Then from Eq.(\ref{eq_def_teta}) we obtain the following asymptotic low-frequency  expression for the decrement in the case of cracks without the waists: 
\begin{equation}
\label{eq23}
\theta_{LF} =6\pi\eta\frac{L^{6} }{h^{3} K} n_{cr} \omega,
\end{equation}
The low-frequency Eq.\,(\ref{eq23}) is obtained  neglecting the liquid compressibility  and agrees with the low-frequency results earlier obtained for cracks without waists \cite{mavko1979wave}. 

To obtain high-frequency expressions (for $\omega>> \omega_{r}$)  we take into account that because of the finite relaxation time, the volume of the replaced  liquid is $\omega/{\omega} _{r}$ smaller than in the low-frequency limit. Thus the
%the resultant asymptotic 
high-frequency expression for the cracks without the waists is 
\begin{equation}
\label{eq24}
\theta_{HF} =6\pi\eta\frac{L^{6} }{h^{3} K} n_{cr} \omega \frac{\omega_{r}^{2}}
{\omega^{2}} =\frac{\pi}{2} n_{cr}L^{3} \frac{ K_{f} }{\alpha K} \frac
{\omega_{r} }{\omega}.
\end{equation}
 For cracks with the waists in a similar way we obtain:
\begin{equation}
\label{eq25}
\widetilde{\theta} _{LF} =6\pi\eta\frac{L^{6} }{\widetilde{h} ^{3} K} \left(
\frac{\widetilde{l} }{L} \right)  \widetilde{n} _{cr} \omega,
\end{equation}
\begin{equation}
\label{eq26}
\widetilde{\theta} _{HF} =6\pi\eta\frac{L^{6} }{\widetilde{h} ^{3} K} \left(
\frac{\widetilde{l} }{L} \right)  \widetilde{n} _{cr} \omega\frac{\widetilde{\omega} _{r}^{2}}
{\omega^{2}} =\frac{\pi}{2}\widetilde{n} _{cr} \frac{L^{3} K_{f} }{\alpha K} 
\frac{\widetilde{\omega} _{r} }{\omega}.
\end{equation}

 Comparing Eqs.(\ref{eq23}), (\ref{eq24}) and  (\ref{eq25}), (\ref{eq26}) with the low- and high-frequency asymptotics of Eqs.(\ref{eq13}) we 
 obtain  the following relaxator-like expressions for the viscous loss at cracks without and with the waists, respectively:
 \begin{equation}
\label{eq27}
 \theta=\frac{\pi}{2} \frac{K_{f} L^{3} }{K\alpha} n_{cr} \frac{\omega
/\omega_{r} }{1+\left(  \omega/\omega_{r} \right)  ^{2} },
\end{equation}
\begin{equation}
\label{eq28}
\widetilde{\theta} =\frac{\pi}{2} \frac{K_{f} L^{3} }{K\alpha} \widetilde{n} _{cr}
\frac{\omega/\widetilde{\omega} _{r} }{1+\left(  \omega/\widetilde{\omega} _{r}
\right)  ^{2} },
\end{equation}
where  $\omega_{r}$ and $\widetilde{\omega}_{r}$ are  given by Eqs.\,(\ref{eq:flow7}).
% and (\ref{eq:flow8}) and the aspect ratio $\alpha$ is intentionally singled out in the pre-factors.  
 
A striking feature of Eqs.\,(\ref{eq27}) and (\ref{eq28}) is that  they have 
% viscous loss at the entire crack with a uniform opening and local loss at the narrow waist in a crack of the same size have 
the same maximum values determined by the characteristic size $L$ of the whole crack. However,  the relaxation frequency in Eq.\,(\ref{eq28}) is strongly shifted from ultrasonic to seismoacoustic frequencies down to $10^0-10^2$ Hz  which were used in field experiments \cite{glinskii2000vibroseismic,bogolyubov2004,Saltykov2006}. 

The next 
%critical for comparison with observations \cite{glinskii2000vibroseismic,bogolyubov2004,Saltykov2006} 
point  is that much smaller mean strains (such as the above discussed tidal strains) can already noticeably affect the opening of the waist. 
%whereas the average opening of the crack and the other parameters remain still practically unchanged. 
This means that due to the variation in the relaxation frequency  $\widetilde{\omega} _{r}$ for the crack with the waist, the position of the relaxation maximum can  noticeably be changed, whereas the height of the relaxation maximum should remain yet practically  unperturbed. Taking into account the equality of the absolute variations $\Delta\widetilde{h} \approx\Delta h $ and the relationship $\varepsilon_{cr} =\varepsilon/\alpha=\varepsilon L/h$ (which we have already used above)
% when estimating the maximum velocity of the fluid flow inside the crack
 we find that the relative variation in the relaxation frequency $\widetilde{\omega} _{r}$ caused by the variation $\Delta \varepsilon$ in the mean strain: 
\begin{equation}
\label{eq29}
\Delta\widetilde{\omega} _{r}^{} /\widetilde{\omega} _{r}^{} =\Delta\varepsilon
\frac {d\widetilde{\omega} _{r}} {d\varepsilon}\frac{1}{\widetilde{\omega} _{r}} \approx
\Delta\varepsilon\frac{3h}{\widetilde{h} }(L/h)=\Delta\varepsilon\frac{3h}{\widetilde{h}} \frac{1}{\alpha}.
\end{equation}
%Equation (\ref{eq29}) shows that even rather weak variations in the mean strain should be able to significantly perturb the position of the relaxation maximum since 
The parameter 
$(3h/\widetilde{h} )/\alpha$ can be very large, for example,  $(3h/\widetilde{h} )/\alpha\sim 10^{6}$ for quite realistic $\alpha\sim 10^{-4}$  and 
$(h/\widetilde{h} )\sim20..30$. 
%, the parameter  $(3h/\widetilde{h} )/\alpha\sim 3\cdot10^{6}$, 
Thus the tidal strains with amplitude $\varepsilon_0 \sim10^{-8} $ can cause the peak-to-peak variation in the 
relaxation frequency  $2\Delta\widetilde{\omega} _{r} /\widetilde{\omega} _{r}$ of several percents and, consequently, comparable variation $2\Delta\widetilde{\theta}/\widetilde{\theta}$ 
of the decrement should be observed at the wings of the relaxation curve. 

Besides, there are known indications that for flows in very narrow gaps (down to nanometer scale), the effective viscosity of the liquid can significantly exceed the viscosity for macroscopic gaps \cite{churaev1971,tas2009capillary}.  This effect can additionally enhance the variations in the dissipation due to variations in the waist opening.  Since small variations in the average strain practically do not yet affect
 the average aspect ratio for the crack, the prefactor  in Eq.\,(\ref{eq28}) remains practically unchanged. Thus the shift of the  relaxation maximum can cause the variation in 
$\Delta\widetilde{\theta}$ of  opposite signs depending on the position of the observation frequency 
$\omega$ relative to the frequency  $\widetilde{\omega}_{r}$ of the relaxation-curve 
maximum similar to the situation for the thermoelastic loss illustrated in Fig.\,\ref{fig:peak-shift}.

Since not all cracks can have the strip-like waists,  it is a key question how many such cracks are required to ensure  near the characteristic frequency $\widetilde{\omega}_{r}$ the dissipation $\theta \sim 10^{-2}-10^{-1}$. Since the maximum of the frequency factor in Eq.\,(\ref{eq28}) is $1/2$ we have to estimate the factor
\begin{equation}
\label{eqdis1}
\widetilde{\theta}_{max} =\frac{\pi}{4} \frac{K_{f} L^{3} }{K\alpha} \widetilde{n} _{cr}=\frac{\pi}{4} \frac{K_{f} }{K\alpha} \widetilde{\epsilon},
\end{equation}
where we singled out the quantity $\widetilde{\epsilon}=L^3\widetilde{n}_{cr}$. The latter is close
to the effective volume of cracks \cite{Connell1974} (i.e., the volume of the circumscribed spheres  independent of the cracks' aspect ratios). For the further 
estimates we will use the well known fact that the
 presence of cracks with the effective volume $\widetilde\epsilon$ results is the reduction of the elastic moduli of the material by a fraction of $a \widetilde\epsilon$ (where factor $a\sim1$ slightly differ for particular moduli \cite{Connell1974}).   Then taking 
 for filling water 
% $\eta=10^{-3}$ Pa$\cdot$s and 
 $K_f=2.25\cdot10^9$ Pa, modulus $K = 3.8\cdot10^{10}$~Pa typical of quartz,  and $\alpha=10^{-3}-10^{-4}$ we obtain $\widetilde{\theta}_{max}\approx(50-500)\widetilde{\epsilon}$. This means that for the effective crack density $\widetilde{\epsilon}=10^{-2}-10^{-3}$, which reduces the elastic moduli also by a fraction of  order $10^{-2}-10^{-3}$, we can already obtain $\widetilde{\theta} \sim 10^{-2}-10^{-1}$ in the vicinity of the relaxation frequency $\widetilde \omega_r$.  On the other hand, it is well known that in real sandstones the modulus reduction due to high-compliant crack-type porosity  typically is on the order of tens of percents \cite{Mavko1994}. This means that even if only a small portion of all cracks (a few percents or even less) has wavy surfaces creating the narrow waists, this portion can already be sufficient to explain both the background value of dissipation observed in 
the seismo-acoustic frequency range $10^0-10^3$ Hz and its extremely high strain sensitivity indicated by  the experimental data  \cite{glinskii2000vibroseismic,bogolyubov2004,Saltykov2006}.
%  as discussed in the introduction. 

Therefore,  the considered modified  mechanism of the squirt-type  dissipation in cracks suggests a plausible  alternative explanation  to the experimentally observed rather high dissipation  in the  seismoacoustic frequency range. In terms of paper \cite{pride2004seismic}, our  mechanism is of purely ``microscopic" type and does not require the presence of  larger-scale ("mesoscale") heterogeneities assumed in the ``double porosity'' and ``patchy saturation'' models   to shift the  frequencies of viscous relaxation towards the seismoacoustic frequency range. Certainly such mechanisms can operate simultaneously. However, among those possibilities, only the above considered modified squirt mechanism is evidently able to ensure sufficiently high strain sensitivity to explain the observations \cite{glinskii2000vibroseismic,bogolyubov2004,Saltykov2006}. 
%Certainly the strain-sensitivity of the conventional squirt loss is  much higher than for ``global'' Biot loss due to fluid flows in rigid pore channels,  however, the increased strain sensitivity of the conventional squirt loss is far not so giant as for the considered modified squirt mechanism. 

To better understand the ratio of contributions of the conventional and the proposed  modified squirt mechanisms, let us   compare the corresponding decrements given by Eqs.(\ref{eq27}) and (\ref{eq28}) in the vicinity of the characteristic relaxation frequency $\omega\sim\widetilde{\omega}_{r}$ of the cracks with the narrow waists. Since $\omega_r \ll \widetilde \omega_r$ for this comparison we
%We recall that a similar comparison was made in the previous section in the case of thermoelastic loss at cracks with and without strip-like contacts (see the schematic elucidation in Fig.\,\ref{contrib}).  The difference is that in the case of the viscous loss, the frequency $\omega\sim\widetilde{\omega}_{r}\ll \omega_{r}$ and thus it corresponds to the low-frequency asymptotic of  expression (\ref{eq27}) for the decrement associated with the conventionally discussed ``global'' viscous dissipation in cracks without the waists.  
take  the low-frequency asymptotic form of  Eq.\,(\ref{eq27}), such that  in the vicinity of $\widetilde{\omega}_{r}$ we obtain
%obtain that in the low-freqeuncy range near the relaxation frequency of the cracks with the waists
\begin{equation}
\label{eq30}
\theta(\omega\sim \widetilde{\omega}_{r})\approx 2\frac{n_{cr}}{\widetilde{n}_{cr}}\frac{\widetilde{\omega}_{r}}{\omega_{r}}\widetilde{\theta}_{max}, 
%\textrm{    where   } \widetilde{\theta}_{0}=\frac{\pi}{2}\frac{K_{f}L^{3}}{K\alpha}\widetilde{n}_{cr}.
\end{equation}
where the difference in the relaxation frequencies $\omega/\widetilde{\omega} _{r}=(h/\widetilde{h})^{3}(\widetilde{l}/L)\gg1$ as follows from Eqs.\,(\ref{eq:flow7}) and (\ref{eq:flow8}).  The factor $(h/\widetilde{h})^{3}(\widetilde{l}/L)$ which can easily reach $10^{2}-10^{4}$, so that quite
  a small 
  %(fractions of a percent) 
  portion of cracks with narrow waists can ensure in the low-frequency range a contribution comparable with or even strongly exceeding the contribution of the majority of other cracks of the same size, but without the 
  waists. For example, a portion $\widetilde{n}_{cr}/ n_{cr}\sim 10^{-2}-10^{-4}$ of cracks with narrow waists is able to ensure extremely high sensitivity of the overall decrement to weak variations in the mean strain. 
  \begin{figure}
\setlength\fboxsep{10pt}
\setlength\fboxrule{5pt}
\includegraphics [width=8cm] {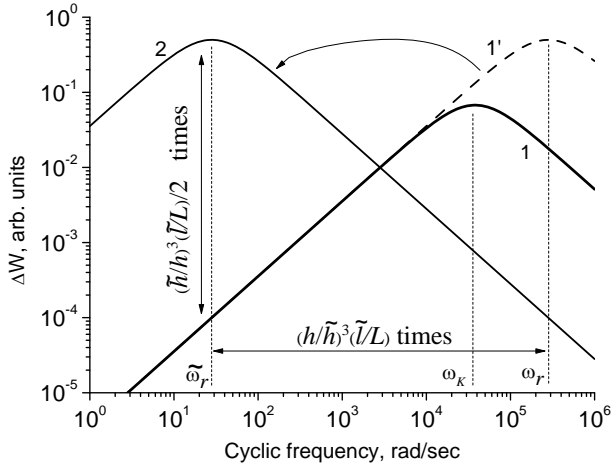}
%\includegraphics [11 8 695 529, width=8cm] {fig-5}
%\fbox{A real figure would go here}
%\source{\cite{earnshaw1842}}
\caption{
Schematically shown relative positions and heights of the viscous relaxation peaks at 
$\omega_{r}$ and ${\omega} _{K}$ for the crack as a whole (curves \emph{1}  and \emph{1'}) and
 the low-frequency peak for a crack of the same size $L$  having an inner narrow waist with 
 the local opening $\widetilde h $ and length $\widetilde l $ (curve \emph{2}). The examples correspond to $(h/\widetilde{h})^3(\widetilde{l}/L)=10^4$ and $\alpha=10^3$. 
}
\label{fig:visc-peaks}
\end{figure} 
 
  Figure \ref{fig:visc-peaks} illustrates the relative positions and heights of the relaxation peaks corresponding to the relaxation frequencies $\widetilde\omega_{r}$ (Eq.\,(\ref{eq:flow7})) for a crack with a narrow waist, and the peaks at frequencies $\omega_{K}$ (Eq.\,(\ref{eq:omega_K})) and $\omega_{r}$ (Eq.\,(\ref{eq:flow8})) for cracks 
  with the same  aspect ratio $\alpha$ and size $L$, but without waists. We emphasize that for 
  $\omega_{K}<\omega_{r}$, the relaxation peak at $\omega_{r}$ actually does not exist and is shown 
  by the dashed line. However, it is shown for convenience of comparison with the 
  low-frequency peak $\widetilde\omega_{r}$, for which the
height is the same as 
 % actual peak at $\omega_K$ and 
  for the would-be peak at $\omega_{r}$.
   %have the same low-frequency asymptotic. 
   The figure demonstrates that the viscous relaxation in cracks with the waists can form 
  extremely strong peaks in the seismo-acoustic frequency range, so that even for small density of such cracks, their contribution can easily account for the typically observed levels of the dissipation in
  this frequency range.  

Concerning the question of averaging over the distribution of real cracks over their parameters, we can put forward very similar arguments as for Eq.\,(\ref{eq5}) in the above considered case of thermoelastic loss. 
%The averaging over the crack parameters should be made similar to the averaging in Eq.\,(\ref{eq5}). Like in Eq.\,(\ref{eq5}) 
Namely, it is reasonable to assume that the distributions over the sizes $L$ and $h$ of the crack as a whole and the distribution over the local parameters $\widetilde{h}$ and $\widetilde{l}$ of the waist (which determine the characteristic relaxation frequency $\widetilde{\omega}_{r}$) are essentially independent and can be factorized.
% like in the case of thermoelastic loss. 
Therefore, the averaging over the characteristic size $L$ of the crack  gives only a numerical factor like in integral  Eq.\,(\ref{eq5}), whereas the averaging over the relaxation frequencies (of, equivalently, the relaxation times) can also be performed independently. Such averaging should not radically change the conclusions obtained for the simplest case of cracks with identical parameters of the inner contacts or waists.  
\section{Conclusion}
The performed analysis of the role of elongated inner contacts and waists in cracks significantly changes the conclusions based on conventionally discussed models  of thermoelastic dissipation at cracks (like \cite{Savage1966, Armstrong1980})
 and viscous squirt loss (like works \cite{Walsh1969,johnston1979-att-mech,mavko1979wave,Murphy1986,pride2004seismic}). 
 %The considered modifications due to wavy asperities on the crack surfaces demonstrate that for thermoelastic dissipation the same crack can provide not only relatively low-frequency (say, below $0.1-1$ Hz) relaxation peak
 %due to the dissipation at the crack as a whole, but a comparable in intensity peak at a much higher
 % frequency determined by the contact width (from $10^2-10^3$~Hz to ultrasonic frequencies as illustrated in Fig.\,\ref{fig:thermo-peaks}). 
  Thus a single larger crack with a strip-like contact can ensure the same thermoelastic dissipation as $10^5-10^6$ small cracks of the size equal to the contact width. 
  %This fact should  significantly affect conclusions on crack distribution over the sizes, in particular, in in the context of nearly frequency-independent decrement typical of many dry rocks. 
 
  For fluid saturated cracks, 
  %the situation is similar, but in this case, in contrast, 
  a rather intense maximum formed by a crack with a narrow waist can ensure the same dissipation in the seismo-acoustic frequency range of $10^2-10^3$~Hz as  a similar crack without the waist would produce  in the ultrasonic range according to conventional squirt-dissipation models (see Fig.\,\ref{fig:visc-peaks}). This can significantly affect some conclusions\cite{pride2004seismic} on insignificant role of local viscous loss at cracks in the seismo-acoustic range.   
 
Probably the most striking feature of the considered modified dissipation mechanisms is their giant strain sensitivity. In this context, it should be clearly understood that each group of cracks with the wavy asperities can exhibit  this giant strain sensitivity   
only in a rather narrow strain range. When the waist becomes either completely closed or widely open, the loss at  
such cracks does not much differ from that at cracks without the asperities. 
Nevertheless, since the parameters of
 real cracks should have rather wide distribution, for a current mean strain, another portion of cracks with such narrow waists or contacts can be ``activated''. This resolves the problem \cite{reasenberg1974precise}  how apparently very soft defects exhibiting giant strain sensitivity can exist under very different 
 mean pressures (in a wide range of depths in field conditions). Their giant effective compliance should be understood as differential. It cannot be directly extrapolated for significantly higher strains $10^{-6}-10^{-4}$ often used in laboratory experiments or typical of tectonic strains, for which the level of the dissipation does not change many times, although for $10^{-8}$ the variations may reach several percents.
 %, although quite significant variations (tens of percents and up to several times) are really observed in some experiments.   

% A similar situation is known for generation of nonlinear distortions of acoustic signals (e.g., harmonics or demodulated components) in granular materials. In a wide range of applied pressures, those signal components are dominated by the weakest fraction of the intergrain contacts. With increasing pressure, previously unloaded contacts become stronger loaded, which strongly decresases their nonlinearity, but simultaneously a new portion of weak contacts is created which  dominate the material nonlinearity. This differential nonlinearity is much stronger than the 
%``smoothed'' nonlinearity estimated via the pressure dependence of the acoustic wave velocity in a wide pressure  range.  
%The following command formats the BiBTeX-generated bibliography by reading in the .bbl file.
%When preparing a TeX document for submission to the JASA, you must paste in the contents
%of that file in place of this command: the Journal requires submission of a single .tex file.
\begin{acknowledgments} We acknowledge the support of the RFBR grants Nos 09-02-91071-CNRS and 11-05-01003.
\end{acknowledgments}

%\bibliography{jasasamp-lev}%

\end{document}